\documentclass[twocolumn,showpacs,prc,aps,preprintnumbers,amsmath,amssymb,superscriptaddress]{revtex4}
\usepackage{amsmath}
\usepackage{graphicx}
\usepackage{subfig}
\usepackage{amssymb}
\usepackage{bm}
\usepackage{color}
\usepackage{makecell}
\usepackage[justification=RaggedRight]{caption}

\begin{document}
\pagestyle{plain}
\title{Effects of tensor and $T=0$ pairing interactions on nuclear $\beta^+$/EC decay}
\author{Feng Wu}
\affiliation{Key Laboratory of Radiation Physics and Technology of Ministry of Education, School of Physics Science and Technology, 
Sichuan University, Chengdu 610065, China}
\affiliation{China Institute of Atomic Energy, Beijing 102413, China}
\author{D. Wu}
\affiliation{Key Laboratory of Radiation Physics and Technology of Ministry of Education, School of Physics Science and Technology, 
Sichuan University, Chengdu 610065, China}
\author{C.L. Bai}
\email[]{bclphy@scu.edu.cn}
\affiliation{Key Laboratory of Radiation Physics and Technology of Ministry of Education, School of Physics Science and Technology, 
Sichuan University, Chengdu 610065, China}
\author{H.Q. Zhang}
\affiliation{China Institute of Atomic Energy, Beijing 102413, China}
\author{X.Z. Zhang}
\affiliation{China Institute of Atomic Energy, Beijing 102413, China}
\author{F.R. Xu}
\affiliation{School of Physics and State Key Laboratory of Nuclear Physics and Technology, \\
Peking University, Beijing 100871, China}

\begin{abstract}
The Hartree-Fock-Bogolyubov (HFB) plus proton-neutron quasiparticle random phase approximation (pnQRPA) approach based on Skyrme 
interaction is applied to study the nuclear $\beta^+$/EC decay for nuclei near the proton magic numbers $Z=$20, 28, 
and 50. With properly selected Skyrme interactions that include the tensor terms, and the $T=0$ pairing interaction, 
the experimental $\beta^+$/EC decay half-lives of these nuclei can be systematically reproduced quite well. 
It is shown that the tensor and $T=0$ pairing interactions play different roles in different nuclei. 
The specific effect is relevant to the configurations that contribute to the decay. 
The attractive and repulsive properties of tensor interaction for the GT state is also studied. 
The present results indicate that it works 
repulsively on the GT states dominated by the configurations from $\pi j_\gtrless$ to $\nu j_\gtrless$ orbits, 
while working attractively for those mainly composed of the configurations from $\pi j_\gtrless$ to $\nu j_\lessgtr$ orbits.
\end{abstract}

\pacs{23.40.-s, 21.10.Tg, 21.60.Jz, 21.30.Fe}

\maketitle
\thispagestyle{plain}
\section{Introduction} \label{introduction}
Nuclear $\beta$ decay plays an important role in nuclear physics. The presumably most important reason 
is the role played by nuclear $\beta^-$ decay in nuclear astrophysics, since the $\beta^-$ decay and 
neutron capture account for the synthesis of elements heavier than iron, and thereby shape the nuclear 
isotopic abundance. Moreover, it also provides insights for nucleon-nucleon interactions.

Theoretically, besides the macroscopic gross theory \cite{Takahashi75}, two main microscopic models, i.e., 
the shell model \cite{Suzuki12,Zhi13} and pnQRPA approach \cite{Moller97, Engel99, Moreno06, Sarriguren11, 
Minato13, Ni14, Ni15, Niu15, Mustonen14, Severyukhin, Mustonen16, Fang16, Niu13, Niu13b, Marketin16}, have been widely applied 
to describe and predict the $\beta$ decay half-lives. In these works, the isoscalar ($T=0$) proton-neutron 
pairing interaction was claimed to have a strong effect on the $\beta$ decay half-lives \cite{Engel99, Niu13, Niu13b}, 
as it shifts the low energy Gamow-Teller (GT) state downward and, therefore, serves to reduce the half-lives. Another 
important interaction that may not be neglected is the tensor interaction, since it can significantly shift 
the low-energy GT states and influence the $\beta^-$ decay half-lives \cite{Bai09, Bai09b, Minato13, Mustonen14}. 
Recently, extensive evaluations for $\beta^-$ decay half-lives have been done in Ref. \cite{Mustonen16} by using 
Skyrme-type interactions with the above two interactions been taken into account. 

In this work, the HFB+pnQRPA approach based on Skyrme interactions is applied 
to study the nuclear $\beta^+$/EC decay for nuclei around proton magic numbers $Z=$20, 28, and 50. Both the effects 
of the tensor and $T=0$ pairing interactions will be studied. 

This paper is organized as follows. In Sec. \ref{formulism}, some necessary details about the HFB+pnQRPA approach and 
essential formulas relevant to the calculation of half-lives for nuclear $\beta^+$/EC decay are presented. In Sec. \ref{parameter}, 
we further make choice of Skyrme parameter sets. Effects of the tensor and $T=0$ pairing interactions are discussed in 
Sec. \ref{tensorT0}. The summary is made in Sec. \ref{summary}.

\section{Formulism}\label{formulism}
In our HFB+pnQRPA calculation, the zero-range Skyrme interactions are employed, and the Skyrme tensor terms read \cite{Skyrme59}:
\begin{eqnarray} \label{tensorf}
V_{T}&=&\dfrac{T}{2}\{[(\mathbf{\sigma _{1}\cdot {k}^{\prime
	})(\sigma _{2}\cdot {k}^{\prime })-\dfrac{1}{3}\left( \sigma
	_{1}\cdot \sigma _{2}\right) {k}^{\prime 2}]\delta(r) }\nonumber\\
&&+\delta (\bf {r})[ (\bf{\sigma _{1}\cdot {k})(\sigma _{2}\cdot
	{k})-\dfrac{1}{3}\left(
	\sigma _{1}\cdot \sigma _{2}\right) {k}^{2}}] \}\nonumber\\
&&+\dfrac{U}{2}\{\left( \sigma _{1}\cdot \bf{k}^{\prime }\right)
\delta (\bf{r}) (\sigma _{2}\cdot \bf{k}) +\left( \sigma _{2}\cdot
\bf{k}^{\prime }\right) \delta (\bf{r})(\sigma _{1}\cdot
{k})\nonumber\\
&&-\dfrac{2}{3}\left[ (\bf{\sigma }_{1}\cdot \bf{\sigma }_{2})
\bf{k}^{\prime }\cdot \delta (\bf{r})\bf{k} \right] \},\label{tensor}
\end{eqnarray}
where $T$ and $U$ are the strengths of triplet-even and triplet-odd tensor terms, respectively, and the operator $\mathbf{k}$$\mathbf{=}$($\mathbf{\nabla_1}$ $-$ $\mathbf{\nabla_2}$)/$\mathbf{2i}$ acts on the right and  
$\mathbf{k^{\prime}}$$\mathbf{=}$$-$($\mathbf{\nabla_1^{\prime}}$ $-$ $\mathbf{\nabla_2^{\prime}}$)/$\mathbf{2i}$ 
acts on the left. In HFB calculation, tensor force only contributes to the spin-orbit potential, which is given by \cite{Colo07, Brink07}
\begin{equation}
U_{so}^{q} = \frac{W_0}{2r} \left( 2 \frac{d \rho_q}{d r} + \frac{d \rho_{q^{\prime}}}{d r} \right) +
\left( \alpha \frac{J_q}{r} + \beta \frac{J_{q^{\prime}}}{r} \right),
\label{uso}
\end{equation}
where $q$ labels either proton or neutron, $\rho$ is the nucleon density, $W_0$ is the strength of the two-body spin-orbit 
interaction, and $J$ is the spin-orbit density. $\alpha$ and $\beta$ receive
the contribution of both the central and tensor terms:
\begin{equation}\label{alphabeta}
\alpha=\alpha_C+\alpha_T,\; \beta=\beta_C+\beta_T,
\end{equation}
with
\begin{eqnarray}
\alpha_C&=&\frac{1}{8}(t_1-t_2)-\frac{1}{8}(t_1 x_1 +t_2 x_2), \nonumber \\ \beta_C&=&-\frac{1}{8}(t_1 x_1 + t_2 x_2), \\
\alpha_T&=&\frac{5}{12}U, \; \beta_T=\frac{5}{24}(T+U).
\end{eqnarray} 

For the pairing force, in the isovector ($T=1$) channel, the zero-range density-dependent surface pairing interaction 
is adopted in both HFB and QRPA calculations as it affects the Fermi transition \cite{Colo05},
and takes the form \cite{Bennaceur05}:
\begin{eqnarray}
V_{T=1}({\bf r_1,r_2})=V_0\frac{1-P_{\sigma}}{2}\left(1-\frac{\rho({\bf r})}{\rho_o}\right)\delta({\bf r_1- r_2}),
\label{T1-pair}
\end{eqnarray}
where ${\bf r = (r_1-r_2)}/2$, $\rho_0$ is taken to be 0.16 fm$^{-3}$, and $P_\sigma$ is the spin exchange operator. 
$V_0$= -470.0 MeV $\cdot$ fm$^3$ is determined by adjusting the empirical pairing gaps of the nuclei studied in 
this work. The $T=0$ pairing interaction is employed in the QRPA calculation and takes the similar form as that in the 
$T=1$ channel \cite{Bai13}:
\begin{equation}
V_{T=0}({\bf r_1,r_2})=fV_0\frac{1+P_{\sigma}}{2}\left(1-\frac{\rho({\bf r})}{\rho_o}\right)\delta({\bf r_1-r_2}),
\label{T0-pair}
\end{equation}
in which $f$ denotes the strength for the $T=0$ pairing interaction relative to that in the $T=1$ channel, and is taken to 
be 1.1 in this work, which is very close to the value recommended in Ref. \cite{Bai14}.

The pnQRPA calculation is based on the ground state of the parent nuclei calculated by solving the HFB equation in coordinate 
space \cite{Dobaczewski84, Bennaceur05}. Under the canonical HFB basis \cite{Dobaczewski96, Terasaki05}, the pnQRPA equation 
takes the form:
\begin{eqnarray} \label{QRPAeq}
\begin{pmatrix} A&B\\-B&-A\end{pmatrix}\begin{pmatrix}X\\Y\end{pmatrix}=E_{QRPA}\begin{pmatrix}X\\Y\end{pmatrix},
\end{eqnarray}
in which $X$ and $Y$ denote the forward and backward amplitudes, $E_{QRPA}$ is the QRPA phonon energy, and 
\begin{eqnarray}
A_{pn,p^\prime n^\prime} &=& E_{pp^\prime}\delta_{nn^\prime}+ E_{nn^\prime}\delta_{pp^\prime}\hspace*{2cm}\notag\\
&&+V_{pn,p^\prime n^\prime}
^{ph}(u_pv_nu_{p^\prime}v_{n^\prime}+v_pu_nv_{p^\prime}u_{n^\prime})\\
&& +V_{pn,p^\prime n^\prime}^{pp} (u_pu_nu_{p^\prime}u_{n^\prime}+v_pv_nv_{p^\prime}v_{n^\prime}),\notag \\
B_{pn,p^\prime n^\prime} &=& V_{pn,p^\prime n^\prime}^{ph}(v_pu_nu_{p^\prime}v_{n^\prime}+
u_pv_nv_{p^\prime}u_{n^\prime})\notag\\
&& - V_{pn,p^\prime n^\prime}^{pp}(u_pu_nv_{p^\prime}v_{n^\prime}+v_pu_nu_{p^\prime}u_{n^\prime}),
\end{eqnarray}
where $p$, $p^{\prime}$, and $n$, $n^{\prime}$ denote the quasiparticle canonical states of proton and neutron, respectively. $u$ and $v$ are the Bogolyubov-Valatin transformation factors. $V^{ph}$ and $V^{pp}$ are the matrix elements for the residual interaction in the particle-hole (ph) and particle-particle (pp) channels, respectively.

After diagonalizing the QRPA matrix, the strength associated with operator 
$\hat{O}_+$ in $t_+$ channel is calculated by
\begin{eqnarray}
B_+^\nu&=&|\sum_{pn}(X_{pn}^\nu v_pu_n+Y_{pn}^\nu u_pv_n)\langle p||\hat{O}_+||n\rangle|^2,
\label{s+}
\end{eqnarray}
in which $\hat{O}_+=\sum\limits_{i}^{A} t_{+ i} \vec{\sigma}_i$ for GT transition, and $\hat{O}_{+}=\sum\limits_{i}^{A} t_{+i}$ for Fermi transition.
For a specific configuration $pn$, we can define a relative transition amplitude:
\begin{eqnarray}\label{ratio}
R_{pn} = [(X_{pn}^\nu v_pu_n+Y_{pn}^\nu u_pv_n)\langle p||
\hat{O}_+||n\rangle]^2/B_+.
\end{eqnarray}
And the value
\begin{eqnarray}\label{norm}
C_{pn}=X_{pn}^2-Y_{pn}^2
\end{eqnarray}
is the normalization factor. In this work, we will use $R_{pn}$ and $C_{pn}$ to evaluate the contribution of a particular $pn$ configuration to a given collective excited state.

The half-life for allowed $\beta$ decay is given by \cite{Moreno06, Niu13}
\begin{equation}
T_{1/2}=\frac{D}{\sum\limits_{\nu}\left[(g_A/g_V)^2_{\text{eff}}B_{GT}(E_{\nu})+B_F(E_{\nu})\right]f(Z,E_{\nu})},
\end{equation}
in which $D=6163.4$ s, and $(g_A/g_V)_{\text{eff}}$=1 is adopted in our calculation. $B_F(E_{\nu})$ and $B_{GT}(E_{\nu})$ 
are the transition strengths for the allowed Fermi (F) and GT transition, which are calculated by the pnQRPA.
In $\beta^+$/EC decay, $f(Z,E_{\nu})$ consists of two parts: positron emission ($f^{\beta^+}$) and electron 
capture ($f^{EC}$). For positron emission, the Fermi integrals $f^{\beta^+}$ is given by
\begin{equation}
f^{\beta^+}(Z,E_m)=\int_{m_e}^{E_m} p_e E_e (E_m-E_e)^2 F_0(Z,E_e) d E_e,
\end{equation}
where $p_e$ and $E_e$ are the emitted positron momentum and energy, $m_e$ is the positron mass, and $F_0(Z,E_e)$ is the Fermi 
function. $E_m$ is the $\beta^+$ decay energy, which can be estimated by
\begin{equation}
E_m=-\Delta_{nH}-m_e-E_x,
\end{equation}
in which $\Delta_{nH}$=0.78227 MeV is the mass difference between the neutron and hydrogen atom. 
$E_x=E_{QRPA}-(\lambda_p-\lambda_n)$ is the QRPA energy with respect to the ground state of the parent nucleus corrected by the difference of the proton and neutron Fermi energy in the parent nucleus. The $\beta^+$ decay energy must be higher than the rest mass of the positron, therefore, the energy threshold for $\beta^+$ decay is $E_x<-\Delta_{nH}-2 m_e$. The decay function $f^{EC}$ for electron capture is given by
\begin{equation}
f^{EC}=\frac{\pi}{2}\sum\limits_x q_x^2 g_x^2 B_x,
\end{equation}
where $x$ denotes the atomic subshell from which the electron is captured, $q$ is the emitted neutrino energy. $g$ and $B$, which 
are calculated by the Dirac-Hartree-Fock approach \cite{Desclaux75, Ankudinov96}, 
are the radial component of the bound-state electron wave function at the nuclear surface, and the correction factor that includes 
the exchange and overlap corrections \cite{Bahcall63, Vatai70}, respectively. And the threshold for EC is $E_x<-\Delta_{nH}$.

\section{Choice of the Skyrme parameter sets}\label{parameter}
This work is devoted to study the $\beta^+$/EC decay half-lives of nuclei around the proton magic numbers $Z=$ 20, 28, and 50, as 
our present HFB+pnQRPA are done in spherical approximation. In the present calculations, the Skyrme parameter T21, T32, T43, and T54 
from T$IJ$ family are employed \cite{Lesinski07}, since they include the central and tensor terms on equal footing, and were checked 
to give reasonable results for the centroid energies of GT and charge-exchange spin-dipole (SD) transitions in $^{90}$Zr and $^{208}$Pb 
as shown in our previous work \cite{Bai11}.

\begin{figure}[!hbt]
	\centering
	\includegraphics[height = 10.0cm, width = 8.0 cm]{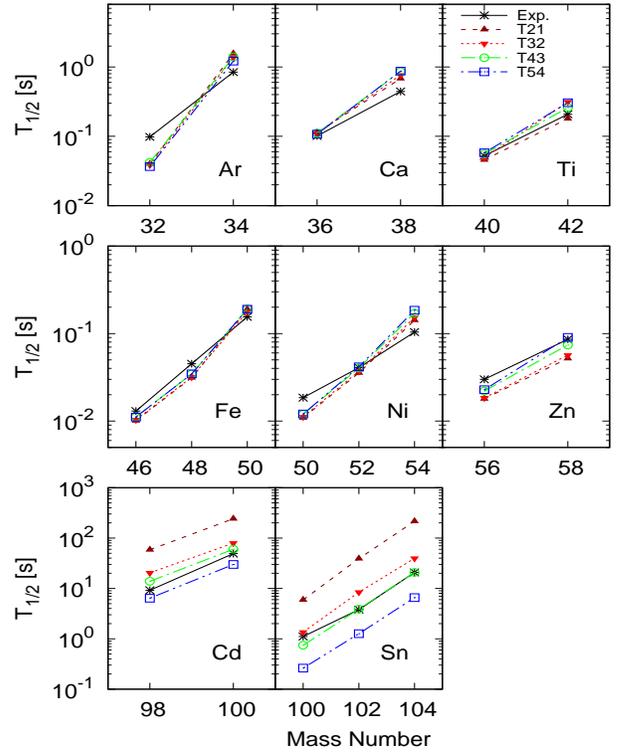}
	\caption{(Color online) Half-lives of Ar, Ca, Ti, Fe, Ni, Zn, Cd, and Sn isotopes calculated by different Skyrme parameter sets, T21, T32, T43, 
		and T54. Both the tensor and $T=0$ pairing interactions are taken into account. The experimental results taken from 
                Ref. \cite{Audi12} are also displayed.}
	\vspace{-1em}
	\label{hlife4t}
\end{figure}

Fig. \ref{hlife4t} shows the calculated half-lives with these four Skyrme parameter sets as well as the experimental results. It is 
seen that the four parameter sets give similar half-lives for Ar, Ca, Ti, Fe, Ni, and Zn isotopes. However, in Cd and Sn isotopes, 
there is great discrepancy among the results of these four Skyrme parameter sets. As we will show in Sec. \ref{tensorT0}, 
the main configurations that contribute to the decay of Ar, Ca, Ti, Fe, Ni, and Zn isotopes are of ($\pi j_\gtrless$ $\rightarrow$ 
$\nu j_\gtrless$) type, while they are of ($\pi j_\gtrless$ $\rightarrow$ $\nu j_\lessgtr$) type for Cd and Sn isotopes. It is known 
that the T$IJ$ parameter sets are fitted by the definition that $\alpha=60(J-2)$ MeV $\cdot$ fm$^5$ and $\beta=60(I-2)$ MeV $\cdot$ 
fm$^5$. The spin-orbit splitting in the four interactions are quite different due to the 
different values of $\alpha$ and $\beta$. The excitation energies of configurations from $\pi j_\gtrless$ to $\nu j_\lessgtr$ orbits
are more sensitive to the change
of spin-orbit splitting as the energy difference between the $\pi j_\gtrless$ and $\nu j_\lessgtr$ orbitals is readily to be amplified
 in these nuclei, in which the spin-orbit splittings for proton and neutron are increased or 
decreased simultaneously. However, the excitation energies of configurations from $\pi j_{\gtrless}$ to $\nu j_{\gtrless}$ orbits are
 quite stable as the energy changes
of $\pi j_\gtrless$ and $\nu j_\gtrless$ orbitals due to
the spin-orbit splitting altering tend to be cancelled. This indicates that reasonable choice of $\alpha$ and $\beta$, and hence
 the tensor interaction is 
essential for theoretical study of $\beta$ decay. Specifically, among these four parameter sets, Fig. \ref{hlife4t} suggests that T32 
and T43 are better choices for the present topic. 

\section{Effects of the tensor and $T=0$ pairing interactions}\label{tensorT0}
\begin{figure}[!hbt]
	\centering
	\includegraphics[height = 10.0cm, width = 8.0 cm]{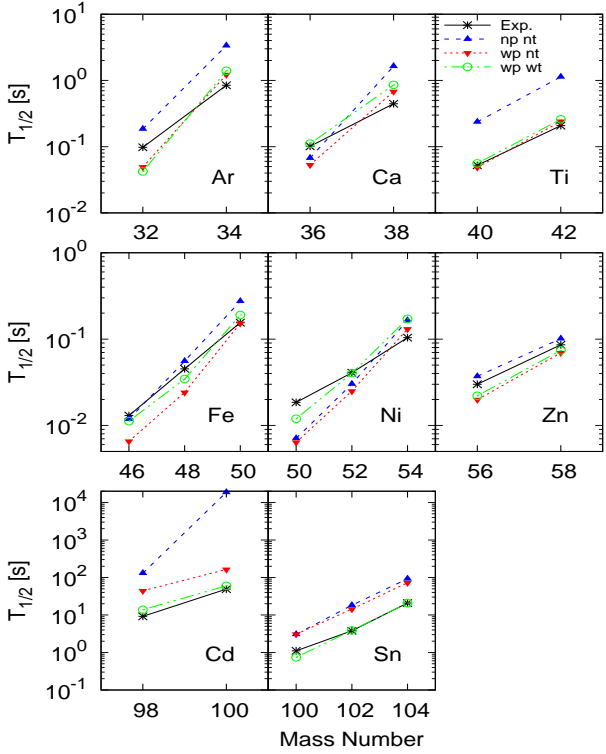}
	\caption{(Color online) $\beta^+$/EC decay half-lives for Ar, Ca ,Ti, Fe, Ni, Zn, Cd, and Sn isotopes calculated by HFB+pnQRPA 
approach with the Skyrme-type parameter set T43. Four kinds of results are displayed: the experimental half-lives 
(denoted as ``Exp.'') taken from Ref. \cite{Audi12}, the results that include neither the $T=0$ pairing interaction nor 
the tensor force (denoted as ``np nt''), the results that only take the $T=0$ pairing interaction into account (denoted 
as ``wp nt''), and the results that take both the $T=0$ pairing and tensor interactions into account (denoted as ``wp wt''). 
The strength for the $T=0$ pairing interaction is 1.1 times of that in the $T=1$ channel.}
	\vspace{-1em}
	\label{hlifet43}
\end{figure}

In this section, the effects of the tensor and $T=0$ pairing interactions on the $\beta^+$/EC decay half-lives
will be studied based on the results calculated with T32 and T43. As T32 produces quite similar results, we only show the results produced by T43 for 
simplicity in the following discussion. The $\beta^+$/EC decay half-lives for Ar, Ca, Ti, 
Fe, Ni, Zn, Cd, and Sn isotopes calculated by HFB+pnQRPA with T43 
are shown in Fig. \ref{hlifet43}. Since both the tensor and $T=0$ pairing interactions may affect the half-lives, three kinds of results are 
presented for comparison: 1) neither the tensor nor the $T=0$ pairing interactions 
are included, and the results are labelled by ``np nt"; 2) only the $T=0$ pairing interaction is taken into account, and the results 
are denoted as ``wp nt"; 3) both the tensor and $T=0$ pairing interactions are included, and the relevant results are labeled by ``wp wt".
 
From Fig. \ref{hlifet43}, one can see that the $T=0$ pairing interaction plays an important role in the evaluation of half-lives for Ar, Ti, and Zn isotopes, while the effect of tensor interaction on the half-lives of these isotopes is quite trivial. The dominant configurations and relevant quantities of the GT states that are important for the evaluation of $\beta^+$/EC decay half-lives of these isotopes are listed in Table \ref{tab1}, in which both the $T=0$ pairing interaction (with $f=1.1$) and the tensor interaction have been taken into account. It is shown that the important GT states of these isotopes are mainly composed of configurations that receive contribution from the pp channel more easily. As a result, the $T=0$ pairing interaction plays a determinant role for the decay rates of these isotopes. Moreover, the good agreement between our theoretical results and the measured half-lives of these nuclei indicates that the present choice for the strength of the $T=0$ pairing interaction is reasonable.
\begin{table}[htbp]
\vspace{-0.5em}
\caption{Important configurations for the $\beta^+$/EC decay of Ar, Ti, and Zn isotopes. The Skyrme interaction T43 is employed, and both the tensor and $T=0$ pairing interactions are taken into account. The abbreviations $R$ and $C$ correspond to the relative transition amplitude and the normalization factor defined in Eqs. (\ref{ratio}) and (\ref{norm}), respectively. The excitation energy $E_x$ is given in unit of MeV.}	
\setlength{\tabcolsep}{2pt}
\renewcommand{\arraystretch}{1.3}
\begin{tabular}{lcccccc}
\toprule
nuclei & $E_x$ & B(GT) & $\pi(v_\pi^2)$ & $\nu(v_\nu^2)$ & R & C \\
\hline
$^{32}$Ar &-11.04  & 0.94 & $2s_{1/2}$(0.888) & $2s_{1/2}$(0.315) & 0.636 & 0.171\\
		  & -10.19 & 1.87 & $2s_{1/2}$(0.888) & $2s_{1/2}$(0.315) & 0.217 & 0.118\\
$^{34}$Ar & -5.99 & 1.52 & $2s_{1/2}$(0.902) & $2s_{1/2}$(0.796) & 0.122 & 0.240\\
		  &       &      & $1d_{3/2}$(0.449) & $1d_{3/2}$(0.125) & 0.299 & 0.557\\
$^{40}$Ti & -8.99 & 4.70 & $1d_{3/2}$(0.921) & $1d_{3/2}$(0.467) & 0.126 & 0.535\\
          &       &      & $1f_{7/2}$(0.230) & $1f_{7/2}$(0.030) & 0.113 & 0.258\\
$^{42}$Ti & -7.31 & 3.88 & $1f_{7/2}$(0.244) & $1f_{7/2}$(0.050) & 0.421 & 0.768\\
$^{56}$Zn & -11.08& 3.61 & $2p_{3/2}$(0.399) & $2p_{3/2}$(0.023) & 0.332 & 0.485\\
$^{58}$Zn & -9.62 & 2.56 & $2p_{3/2}$(0.375) & $2p_{3/2}$(0.000) & 0.846 & 0.872\\
\botrule	  
\end{tabular}
\label{tab1}
\end{table}

On the contrary, for $^{50}$Ni and Sn isotopes, the tensor interaction plays a dominant role while the 
impact of $T=0$ pairing interaction is quite small. 
As tabulated in Table \ref{tab2}, the configurations that are important for the decay can more readily 
receive contribution from the 
ph channel. Therefore, interactions in the ph channel, specially the tensor interaction, play an important 
role in the evaluation for half-lives of 
these nuclei.
\begin{table}[htbp]
\caption{Same as Table \ref{tab1}, but for $^{50}$Ni and Sn isotopes.}	
\setlength{\tabcolsep}{2pt}
\renewcommand{\arraystretch}{1.3}
\begin{tabular}{lcccccc}
\toprule
nuclei & $E_x$ & B(GT) & $\pi(v_\pi^2)$ & $\nu(v_\nu^2)$ & R & C \\
\hline
$^{50}$Ni &-12.26 & 5.03 & $1f_{7/2}$(1.000) & $1f_{7/2}$(0.232) & 1.557 & 0.998\\
$^{100}$Sn & -5.33&18.13 & $1g_{9/2}$(1.000) & $1g_{7/2}$(0.000) & 0.949 & 0.978\\
$^{102}$Sn & -4.05& 16.00& $1g_{9/2}$(1.000) & $1g_{7/2}$(0.064) & 1.002 & 0.968\\
$^{104}$Sn & -3.25& 8.83 & $1g_{9/2}$(1.000) & $1g_{7/2}$(0.129) & 0.869 & 0.496\\
           & -3.04& 5.63 & $1g_{9/2}$(1.000) & $1g_{7/2}$(0.129) & 1.317 & 0.480\\
\botrule	  
\end{tabular}
\label{tab2}
\end{table}

Apart from the above two situations, there are also many nuclei in which both the tensor and $T=0$ pairing 
interactions have non-negligible impact. As listed in Table \ref{tab3}, the main configurations that are 
important for the decay can conveniently receive contributions from both the pp and ph channel. Thus, both 
the $T=0$ pairing and tensor interactions can obviously influence the calculated half-lives of these nuclei. 
Specially, by comparing the results of Fe or Ni isotopes, one can find that as more neutrons occupy in the 
$\nu 1f_{7/2}$ orbital, the effect of $T=0$ pairing interaction relative to that of the tensor force gradually 
become larger. This indicates that the $T=0$ and tensor interactions compete with each other in some nuclei 
whose decay are mainly built from some specific configurations.

Furthermore, as shown in Fig. \ref{hlifet43}, the attractive $T=0$ pairing interaction universally helps to 
reduce the half-lives as it can shift the excited states downward to low energy region, which has already been 
demonstrated in many works \cite{Engel99, Niu13, Niu13b}. The effects of the tensor interaction on the half-lives are 
not so monotonous: it reduces the calculated half-lives in Cd and Sn isotopes, while increasing the calculated 
half-lives in Ca, Fe, and Ni isotopes. This indicates that tensor interaction can work both attractively and 
repulsively. Moreover, it is seen that the inclusion of tensor interaction generally helps to improve the theoretical 
results: it increases the half-lives when the $T=0$ pairing interaction reduces it too much, while decreases the 
half-lives when the $T=0$ pairing interaction does not reduce it sufficiently.
\begin{table}[htbp]
\vspace{-0.5em}
\caption{Same as Table \ref{tab1}, but for Ca, Fe, $^{52}$Ni, $^{54}$Ni, and Cd isotopes.}
\setlength{\tabcolsep}{2pt}
\renewcommand{\arraystretch}{1.3}
\begin{tabular}{lcccccc}
\toprule
nuclei & $E_x$ & B(GT) & $\pi(v_\pi^2)$ & $\nu(v_\nu^2)$ & R & C \\
\hline
$^{36}$Ca & -9.03 & 1.57 & $2s_{1/2}$(1.000) & $2s_{1/2}$(0.841) & 0.211 & 0.353\\
$^{38}$Ca & -6.48 & 1.58 & $1d_{3/2}$(0.806) & $1d_{3/2}$(0.443) & 0.526 & 0.879\\
$^{46}$Fe &-11.56 & 7.53 & $1f_{7/2}$(0.667) & $1f_{7/2}$(0.000) & 0.861 & 0.950\\
$^{48}$Fe & -9.86 & 5.44 & $1f_{7/2}$(0.681) & $1f_{7/2}$(0.232) & 0.787 & 0.906\\
$^{50}$Fe & -7.87 & 2.70 & $1f_{7/2}$(0.695) & $1f_{7/2}$(0.463) & 1.030 & 0.972\\
$^{52}$Ni &-10.48 & 3.05 & $1f_{7/2}$(1.000) & $1f_{7/2}$(0.467) & 1.811 & 1.012\\
$^{54}$Ni & -8.79 & 1.37 & $1f_{7/2}$(1.000) & $1f_{7/2}$(0.715) & 2.190 & 1.023\\
$^{98}$Cd & -3.22 & 6.66 & $1g_{9/2}$(0.729) & $1g_{7/2}$(0.000) & 0.926 & 0.481\\
$^{100}$Cd & -3.32& 3.26 & $1g_{9/2}$(0.731) & $1g_{7/2}$(0.055) & 0.498 & 0.135\\
\botrule
\end{tabular}
\label{tab3}
\end{table}
\begin{table}[htbp]
	\vspace{-0.5em}
	\caption{The excitation energies ($E_x$) of the GT states that play the most important 
role in the evaluation of $\beta^+$/EC decay half-lives of Ca, Fe, Ni, Cd, and Sn isotopes, in unit of MeV. 
The HFB+QRPA calculations are done with and without tensor interaction, and the results are labelled by $wt$
and $nt$, respectively. The $T=0$ pairing interaction is included with $f$ = 1.1.}
	\setlength{\tabcolsep}{7pt}
	\renewcommand{\arraystretch}{1.3}
	\begin{tabular}{lccclccc}
		\toprule
		nuclei & $E_{nt}$ & $E_{wt}$ &nuclei & $E_{nt}$ & $E_{wt}$ \\
		\hline
		$^{36}$Ca & -10.95& -9.03 & $^{54}$Ni & -9.41 & -8.79 \\
		$^{38}$Ca & -6.92 & -6.48 & $^{98}$Cd & -3.56 & -4.19\\
		$^{46}$Fe & -12.86& -11.56& $^{100}$Cd& -3.11 & -3.32\\
		$^{48}$Fe & -10.72& -9.86& $^{100}$Sn& -4.32 & -5.33\\
		$^{50}$Fe & -8.42 & -7.87& $^{102}$Sn & -3.15 & -4.05\\
		$^{50}$Ni & -13.80& -12.26 &$^{104}$Sn & -2.30 & -3.25 \\
		$^{52}$Ni & -11.58& -10.48 &          &      &       &  \\
		\botrule
	\end{tabular}
	\label{tab4}
\end{table}

The excitation energies ($E_x$) of the GT states important for the $\beta^+$/EC decay of Ca, Fe, Ni, Cd, and Sn isotopes calculated 
with and without the tensor interaction are listed in Table \ref{tab4}. From Table \ref{tab4}, one may readily see that the tensor 
interaction shifts the low energy GT states of Ca, Fe, and Ni isotopes upward, while shifts the low energy GT states of Cd and Sn isotopes 
downward. As shown in Tables \ref{tab2} and \ref{tab3}, the main configurations for $\beta^+$/EC decay in Ca, Fe, and Ni isotopes 
are of ($\pi j_\gtrless$ $\rightarrow$ $\nu j_\gtrless$) type, while they are of ($\pi j_\gtrless$ $\rightarrow$ $\nu j_\lessgtr$) type 
for Cd and Sn isotopes. Our present results indicate that the tensor 
interaction works repulsively in the GT states dominated by the configurations from $\pi j_\gtrless$ to $\nu j_\gtrless$ orbits, 
and works attractively for those dominated by the configurations from $\pi j_\gtrless$ to $\nu j_\lessgtr$ orbits. 
In fact, it was reported in Ref. \cite{Otsuka05} that the tensor interaction is repulsive between $\pi j_\gtrless$ and 
$\nu j_\gtrless$ orbits, while it is attractive between $\pi j_\gtrless$ 
and $\nu j_\lessgtr$ orbits. This effect of tensor interaction was also easily shown to be valid in mean field 
calculation by using Skyrme force with the strength of tensor terms chosen
to make negative $\alpha$ and positive $\beta$ \cite{Colo07, Brink07}. Our results indicate that this property of tensor 
interaction might still be valid in the GT state for the presently studied nuclei, but only the values of $\beta$ to be positive
are required in the present study.

\section{SUMMARY}\label{summary}
The HFB+pnQRPA approach based on Skyrme-type interaction has been applied to study the nuclear $\beta^+$/EC decay for nuclei around the 
proton magic numbers $Z=$20, 28, 50. With properly selected $T=0$ pairing interaction and Skyrme interaction that includes the tensor 
terms, the experimental half-lives of these nuclei can be well reproduced. 

It is shown that the tensor and $T=0$ pairing interactions play different roles in different nuclei. The specific effect is relevant to 
the configurations that contribute to the decay. For Ar, Ti, Zn isotopes, the $T=0$ pairing interaction plays a dominant role, while 
for $^{50}$Ni, and Sn isotopes, the tensor force is much more important. Besides, for Ca, Fe, $^{52}$Ni, $^{54}$Ni, and Cd isotopes, 
both the $T=0$ pairing and tensor interactions may make the differences. Moreover, our present studies indicate that the tensor inteaction
 works repulsively for the GT states dominated by the configurations from $\pi j_\gtrless$ to $\nu j_\gtrless$ orbitals, 
while working attractively for those maily composed of the configurations from $\pi j_\gtrless$ to $\nu j_\lessgtr$ orbitals.
This property of the tensor interaction is consistent with the one in which the tensor interaction works on the single-particle states,
as shown in shell model and mean field calculations.

\begin{acknowledgments}
This work is supported by the National Natural Science Foundation of China under Grant Nos. 11575120 and 11375266.
\end{acknowledgments}

\end{document}